\documentclass[aps,prb,reprint,superscriptaddress,floatfix,amsmath,showpacs]{revtex4-1}

\usepackage{graphicx}% Include figure files
\usepackage{times}% Times New Roman
\usepackage{textcomp}
\usepackage{amsmath}

\begin{document}

\title{
%Variable
Damping and coherence in a high-density magnon gas}

\author{S. Sch\"{a}fer}
\affiliation{Fachbereich Physik and Forschungszentrum OPTIMAS, Technische Universit\"{a}t Kaiserslautern, 67663
Kaiserslautern, Germany}

\author{V. Kegel}
\affiliation{Fachbereich Physik and Forschungszentrum OPTIMAS, Technische Universit\"{a}t Kaiserslautern, 67663
Kaiserslautern, Germany}

\author{A. A. Serga}
\affiliation{Fachbereich Physik and Forschungszentrum OPTIMAS, Technische Universit\"{a}t Kaiserslautern, 67663
Kaiserslautern, Germany}

\author{B. Hillebrands}
\affiliation{Fachbereich Physik and Forschungszentrum OPTIMAS, Technische Universit\"{a}t Kaiserslautern, 67663
Kaiserslautern, Germany}

\author{M. P. Kostylev}
\affiliation{School of Physics, M013, University of Western Australia, Crawley, WA 6009, Australia}

\date{\today}% It is always \today, today, but any date may be explicitly specified

\begin{abstract}
We report on the fast relaxation behavior of a high-density magnon gas created by a parametric amplification process. The magnon gas is probed using the technique of spin-wave packet recovery by parallel parametric pumping. Experimental results show a damping behavior which is in disagreement with both the standard model of exponential decay and with earlier observations of non-linear damping. In particular, the inherent magnon damping is found to depend upon the presence of the parametric pumping field. A phenomenological model which accounts for the dephasing of the earlier injected magnons is in good agreement with the experimental data.
\end{abstract}

% PACS, the Physics and Astronomy Classification Scheme.
%75.30.Ds -- Magnetic properties - of magnetically ordered materials, - spin waves; Magnons; Spin waves
%76.50.+g -- Ferromagnetic, antiferromagnetic, and ferrimagnetic resonances; spin-wave resonance
%76.20.+q -- General theory of resonances and relaxations
%85.70.Ge -- Ferrite devices, Garnet devices


\maketitle

\textbf{I. Introduction}

Several decades after the first phenomenological descriptions, the nature of spin-wave dissipation is still a subject of intense scientific research \cite{Suhl1998,Gilbert2004,Rezende2001,Azevedo2000, Dobin2003, Gilmore2010}. An understanding of spin-wave damping mechanisms is crucial both for the furtherance of the field of fundamental spin dynamics \cite{Demokritov2006} and practical applications in novel magnetic memory, microwave, and spin-wave logic devices \cite{Khitun2010}. Recent developments in the study of energy transfer in magnon systems, including the discovery of Bose-Einstein condensation of magnons \cite{Dziapko}, have attracted particular interest to the problem of nonlinear damping and decoherence of densely populated magnon states, created by the effect of parametric pumping \cite{Suhl1957, White1963}.

Up to now there has been no access to the information on relaxation of highly populated short-wavelength magnon states as these spin waves are too short to be directly detected by conventional microwave and optical methods. Only measurements of thresholds of parametric instability provide us with information about the decay characteristics of these magnons at low population level. However, even in this case the magnons are influenced by an external microwave pump field. To overcome this obstacle and to access the freely evolving highly populated magnon groups, we used the recently developed technique of spin-wave signal recovery \cite{Serga2007}.

In this work we demonstrate that free evolution of a dense group of parametrically injected magnons can be described neither by its monotonic exponential relaxation nor by possible nonlinear decay. The phase decoherence of the frequency smeared magnon group should be taken into account in order to understand the experimental data. We chose a single crystal yttrium-iron-garnet (YIG) ferrimagnetic film as a test object in our experiments. Due to its extremely low natural magnetic damping \cite{Cherepanov1993} this material is widely used in works on magnon gases and condensates \cite{Demokritov2006}. Furthermore, currently one observes a clear revival of interest to YIG films due to the recent discovery of the spin Hall \cite{Kajiwara2010} and spin Seebeck \cite{Uchida2010} effects in Platinum/YIG bi-layers. Moreover, submicron-thick polycrystalline YIG films are now promising as candidates for novel microwave devices \cite{Grishin}, and conventional single-crystal micron-thick films have been recently used to demonstate microwave encoding \cite{Ordonez}.

\begin{figure}[]
\includegraphics[width=0.85\columnwidth]{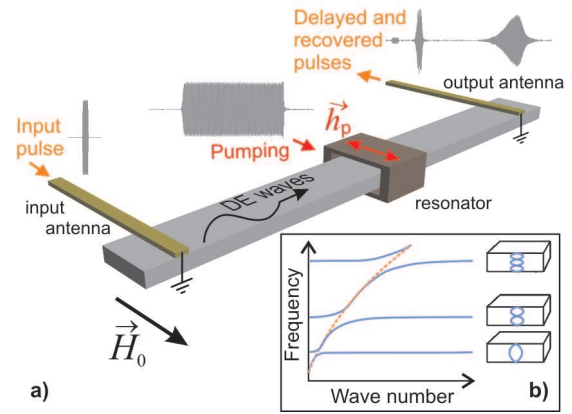}
\caption{\label{Figure1} (Color online) a) Experimental setup, consisting of a single-crystal YIG waveguide, input and output antennas and a dielectric resonator for the application of the pumping microwave field. The waveforms are the shapes of the input, pumping and restored pulses as seen on the oscilloscope.
b) Section of the spin-wave dispersion spectrum for a thin magnetic waveguide schematically showing the hybridization of Damon-Eshbach (dashed line) wave with standing spin wave modes.
}
\end{figure}

\vspace{0.5cm}
\textbf{II Experiment}

\begin{figure}
\includegraphics[width=0.95\columnwidth]{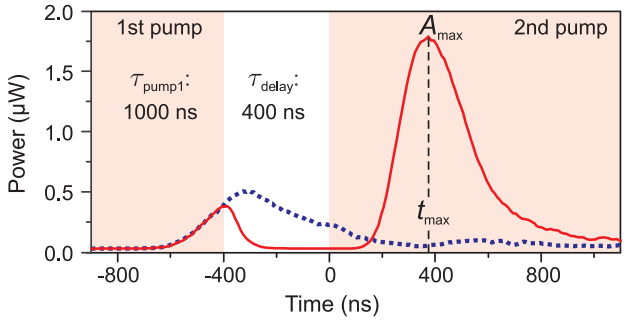}
\caption{ \label{Figure2} (Color online) Spin-wave power envelope as experimentally observed at the oscilloscope. The light red areas schematically show the time intervals when the pump is on. The continuous red line represents the results of the two pump-pulse recovery. The first pump pulse is switched off before the recovered signal saturates. After a delay time $\tau_{\mathrm{delay}}$ the pumping is switched on a second time the second recovered signal is observed. The dashed blue line represents the experimental result for the pumping not being interrupted.
}
\end{figure}

A 100~ns-long traveling spin-wave packet $(\tau_\mathrm{input}=100$~ns) is excited at a frequency of 7~GHz by a microstrip input antenna in a 1.5~mm-wide and 5~{\textmu m}-thick YIG stripe. An external magnetic field $\vec{H}_0$ of 143.64~kA/m (1805~Oe) is applied in the plane of the stripe, perpendicular to the spin-wave propagation (Fig.~\ref{Figure1}a), i.e. in the Damon-Eshbach (DE) geometry \cite{Damon1960}. The excited spin-wave packet traverses the magnetic sample, is detected by the output antenna placed 12~mm apart from the input one, amplified, and observed with an oscilloscope.  After 450~ns since the passage by the wave packet of the central area of the YIG stripe ($\tau_0=450$~ns), a pumping microwave field $\vec{h}_\mathrm{p}$ at 14~GHz produced by a dielectric resonator is applied parallel to the bias magnetic field. As a result of the pump action, an additional ``recovered'' pulse is observed at the output antenna at a frequency of 7~GHz (see output waveform in Fig.~\ref{Figure1}a and dashed line in Fig.~\ref{Figure2}). Since the DE packet has already left the area of parametric interaction, this is not the result of the direct amplification of the signal pulse \cite{Kalinikos1996, Kalinikos1997}. Rather, the parametric pumping acts on the standing spin-wave (SSW) modes (see Fig.~\ref{Figure1}b) which are excited across the film thickness by the traveling wave packet through a two-magnon scattering mechanism. These modes are then amplified by parametric pumping and are scattered back to form a new traveling DE wave, which is finally detected at an output antenna in a form of the ``recovered'' pulse (Fig.~\ref{Figure1}a).

Unlike in previous studies \cite{Serga2007, Schafer2008, Chumak2009}, in the experiment reported here, two consecutive parametric pump pulses are used rather than one. The response of the magnonic system to the second pulse is utilized as a probing tool for extracting information about the free relaxation behavior of parametrically excited magnons during the pump free pause. Therefore, we cut off the first recovered pulse by stopping the parametric pumping. After a certain delay time $\tau_\mathrm{delay}$, a second pumping pulse is applied and a second recovered pulse is observed (see continuous line in Fig.~\ref{Figure2}). We register the peak amplitude $A_\mathrm{max}$ and the arrival time $t_\mathrm{max}$ of the second recovered pulse.

\begin{figure}
\includegraphics[width=1\columnwidth]{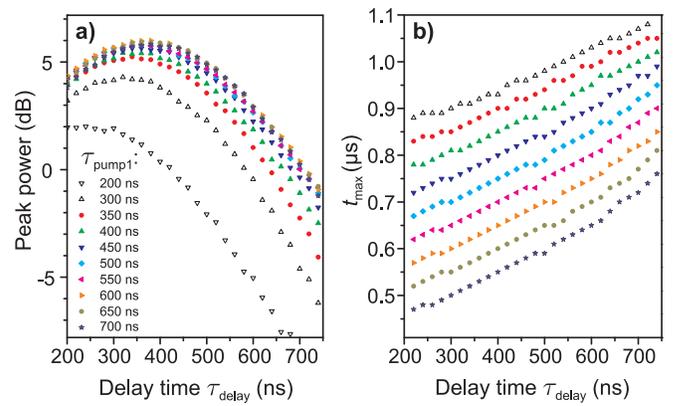}
\caption{ \label{Figure3} (Color online) a) The amplitudes of the second pulse shown in Fig.~\ref{Figure2}a are extracted and plotted against $\tau_{\mathrm{delay}}$ for different durations $\tau_{\mathrm{pump1}}$ of the first pumping pulse. The amplitudes are normalized to the maximum observable amplitude using only one long pumping pulse.
b) The time $t_\mathrm{max}$ at which the peak amplitude of the second pulse in Fig.~\ref{Figure2}a is observed as a function of $\tau_{\mathrm{delay}}$.
}
\end{figure}

As one sees from Fig.~\ref{Figure3}, the behavior of the second recovered pulse strongly depends on the duration of the first pumping pulse $\tau_{\mathrm{pump1}}$ for small $\tau_{\mathrm{pump1}}\leq 450$~ns as well as on the delay time $\tau_{\mathrm{delay}}$ between two consecutive pumping pulses.  It is remarkable that the latter dependence is non-monotonic: with an increase in $\tau_{\mathrm{delay}}$ the second recovered pulse first grows, reaches a maximum, and then decreases in intensity (Fig.~\ref{Figure3}a). In particular, one observes a gain in the amplitude of the second recovered pulse compared to the experiment in which the pumping is uninterrupted ($\tau_{\mathrm{delay}}\rightarrow0$ in Fig.~\ref{Figure3}a).
\vspace{0.5cm}

\textbf{III Theoretical model}

One easily finds that neither the initial increase in the amplitude nor the subsequent non-exponential decrease (note the logarithmic scale) in the peak amplitude can be explained in the framework of the standard approach based on association of a specific relaxation time with each decay process. For this reason we make an attempt to employ a more detailed model. The model we use is an extension of the theory in Ref.\cite{Chumak2009}. The latter succeeded in describing the single-pulse pumping. Treating  a two-pulse pumping with a significant length of the pause between the two pulses requires considering dephasing within the parametrically excited wave packet during the pause. Thus, in contrast to Ref.\cite{Chumak2009}, one cannot use ensemble-averaged equations for magnon densities and has to consider each magnon group as consisting of a number of pairs of waves with different eigenfrequencies.

We start the description of the developed theory  by noting that the recovered pulse has a pulse-like shape. This suggests that competition of two magnon groups for energy provided by the parametric pumping process \cite{L'vov1994} takes place in the magnon system, since a pulse-like shape of the observed signal is a typical result of this competition. This process can be explained in the following way. The parametric amplification is frequency selective: only magnon groups with eigenfrequencies $\omega_k$ which lie inside the narrow frequency band $\omega_\mathrm{p}/2-\nu<\omega(k)<\omega_\mathrm{p}/2+\nu$ whose width is equal to the parametric amplification gain $\nu=h_\mathrm{p} V$ are amplified ($V$ is the parametric coupling coefficient). They are driven at half the frequency $\omega_\mathrm{p}/2$ of the applied microwave field. Importantly, there is no restriction on  the magnitude and the direction of the magnon wave vector: the process allows creation of magnons with arbitrary wave vectors, provided their frequencies are inside this frequency band. Thus, given the large number of frequency-degenerate magnon dispersion branches, magnon groups over a large range of wave vectors are amplified. This includes externally excited oscillations, hereafter called the \textit{signal group}, as well as thermally activated magnons with wave vectors up to $10^5$~$\mathrm{cm}^{-1}$. The latter magnons, which are considerably decoupled from structural defects of the YIG film because of their short wavelength, experience the lowest two-magnon decay and consequently show the highest parametric gain~\cite{Zakharov1975}. Hence, we refer to this dominantly amplified magnon group as the \textit{dominant group}.

As the phase shift between the external pumping field $\vec{h}_\mathrm{p}$ and a microwave magnetic field induced by the parametrically pumped magnons (``internal pumping'') increases with increase in the total magnon density, the resulting effective pumping field inside the sample consecutively diminishes \cite{Chumak2009}. At some point in time the total magnon density reaches a critical threshold level $A_\mathrm{cr}$ at which the effective pumping becomes so small that it is able to further support just one magnon group, the one which has the lowest decay rate and the highest efficiency of parametric interaction: the dominant group. This leads to the suppression of the signal group, whilst the dominant group reaches saturation at $A_\mathrm{cr}$.  From this time on the system stays in a quasi-equilibrium state wherein the magnon density of the dominant group is saturated. The signal group is exposed to an effective pumping below the generation threshold~\cite{Suhl1957} and thus should decay with time to the thermal level. As a result of this process, the signal group, and consequently the measured signal, acquire a pulse-like shape

To obtain a quantitative description of the magnetic dynamics for the present case of two pump pulses, we perform numerical simulations based on the nonlinear model of the magnon group competition. We assume that the signal group represents a set of spin-wave oscillations (standing-wave modes) $C_{\omega}^\mathrm{s}$ with resonance frequencies $\omega$ uniformly distributed across the frequency band $\delta \omega = \pm 1.5$~MHz around the half-pump frequency $\omega_\mathrm{p}/2$. This frequency band corresponds to the spectral width of the dipole-exchange gaps in the spin-wave dispersion (Fig.~1b). Since the recovered pulse is only observed in these gaps, only magnons within this frequency band should contribute to the recovery process.

The standing-wave modes are excited through 2-magnon scattering processes by the  traveling-wave pulse on its way across the film. For simplicity we assume that the initial phase of magnetization precession for all the modes is the same. The output traveling-wave pulse is formed due to 2-magnon back-scattering and thus represents an exact time replica of the dynamics of the signal group. Therefore it is sufficient to consider only the evolution of two \textit{immobile} magnon groups: the signal and the dominant ones (The dominant group consists of magnons with complex amplitudes $C_{\omega}^\mathrm{d}$.)

The general dynamic equations are as follows:
\begin{eqnarray}
\left[\partial /\partial t + \Gamma_\mathrm{s} + i(\tilde{\omega}^\mathrm{s} - \omega_\mathrm{p}/2)\right] C_{\omega}^\mathrm{s}-iP_{\omega}^\mathrm{s}C_{\omega}^\mathrm{s \ast} & = & 0 , \\
\left[\partial /\partial t + \Gamma_\mathrm{d} + i(\tilde{\omega}^\mathrm{d} - \omega_\mathrm{p}/2)\right] C_{\omega}^\mathrm{d}-iP_{\omega}^\mathrm{d}C_{\omega}^\mathrm{d \ast} & = & 0 ,
\end{eqnarray}
where the stars denote complex conjugation, $\Gamma_s$ and $\Gamma_d$ are the relaxation parameters for the signal and the dominant group respectively,
\begin{eqnarray}
\tilde{\omega}^\mathrm{s} & = & \omega+\Delta \omega^\mathrm{ss}+\Delta \omega^\mathrm{sd} , \\
\tilde{\omega}^\mathrm{d} & = & \omega+\Delta \omega^\mathrm{dd}+\Delta \omega^\mathrm{ds}
\end{eqnarray}
are the eigenfrequencies of standing-wave oscillations with an account of nonlinear frequency shift for both signal and dominant groups respectively, and
\begin{eqnarray}
%P_{\omega}=\nu+S\left(\sum_{\omega'}C_{\omega'}^{(s)}^2 +\sum_{\omega'}C_{\omega'}^{(d)}^2\right)
P^\mathrm{s}_{\omega}=\nu+P^\mathrm{ss}+P^\mathrm{sd} , \\
P^\mathrm{d}_{\omega}=\nu+P^\mathrm{ds}+P^\mathrm{dd}
\end{eqnarray}
is the effective pumping for the signal and the dominant groups respectively. In Eqs.~(3-6) and below the upper indices ``s'' and ``d'' denote the signal and the dominant group respectively. Double indices of type $\alpha \beta$ denote the action of the group $\beta$ on the group $\alpha$.
If both indices are the same the respective coefficient or magnitude describes self-action.

In our calculations we make a number of important simplifications which allow us to considerably decrease the computation time. First we use a usual assumption \cite{Chumak2009} that the nonlinear frequency shift is not important for the dynamics of the dominant group ($\Delta \omega^\mathrm{dd}=0$). Second, we do not need to take into account the contribution to the total nonlinear frequency shift by the signal group ($\Delta \omega^\mathrm{ss}=\Delta \omega^\mathrm{ds}=0$), since its amplitude is considerably smaller than the amplitude of the dominant group at the time, when the effective pumping saturates. Thus, the only term we have to account for is
\begin{equation}
\Delta \omega^\mathrm{sd}=T^\mathrm{sd} \!\!\!\! \sum_{\omega', \omega'', \omega'''} \!\!\!\!
C^\mathrm{d}_{\omega'} C^\mathrm{d \ast}_{\omega''} C^\mathrm{s}_{\omega'''} \, \delta(\omega'-\omega''+\omega'''-\omega),
\end{equation}
where $\delta$ denotes the Kronecker Delta, and $T^\mathrm{sd}$ is the respective nonlinear coefficient. In the following we will use a short hand notation $T^\mathrm{sd}=T$.

Similar considerations apply to the contributions to the total pumping which consists of the external pumping ($\nu$)
and the internal pumping given by the remainder of the terms in Eqs.~(5, 6). We obtain: $P^\mathrm{ss} = P^\mathrm{ds} = 0$,
\begin{eqnarray}
P^\mathrm{sd} & = & S^\mathrm{sd} \!\!\!\! \sum_{\omega', \omega'', \omega'''} \!\!\!\! C^\mathrm{d}_{\omega'}C^\mathrm{d}_{\omega''}C^\mathrm{s \ast}_{\omega'''} \, \delta(\omega'-\omega''+\omega'''-\omega) , \\
P^\mathrm{dd} & = & S^\mathrm{dd} \!\!\!\! \sum_{\omega', \omega'', \omega'''} \!\!\!\! C^\mathrm{d}_{\omega'}C^\mathrm{d}_{\omega''}C^\mathrm{d \ast}_{\omega'''} \, \delta(\omega'-\omega''+\omega'''-\omega),
\end{eqnarray}
where $S^{\alpha \beta}$ are the respective nonlinear $S$-coefficients \cite{Chumak2009, L'vov1994}.

Test numerical calculations show that calculation results do not qualitatively change with the variation in the difference $S^\mathrm{dd}-S^\mathrm{sd}$ in reasonable limits. Therefore, to minimize the number of degrees of freedom we set $S^\mathrm{dd}=S^\mathrm{sd} \equiv S$ and, as previously, we assume that $S>0$ \cite{Chumak2009}.

The last simplification we make is removing the frequency-mixing terms from Eqs.~(7-9). This simplification does
not lead to a qualitative change in the results of our numerical calculations, but enormously decreases the computation time. It reduces Eq.~(4) to a simple formula for the effective pumping:
\begin{equation}
P^\mathrm{s}_{\omega}=P^\mathrm{d}_{\omega} \equiv P_{\omega} = \nu + S \sum_{\omega'} \left( C_{\omega'}^\mathrm{d} \right)^2.
\end{equation}

For the pump-free period the same equations (1-10) are valid, one just assumes $\nu = 0$, so that only the internal pumping is on during the pause.

\begin{figure}
\includegraphics[width=0.95\columnwidth]{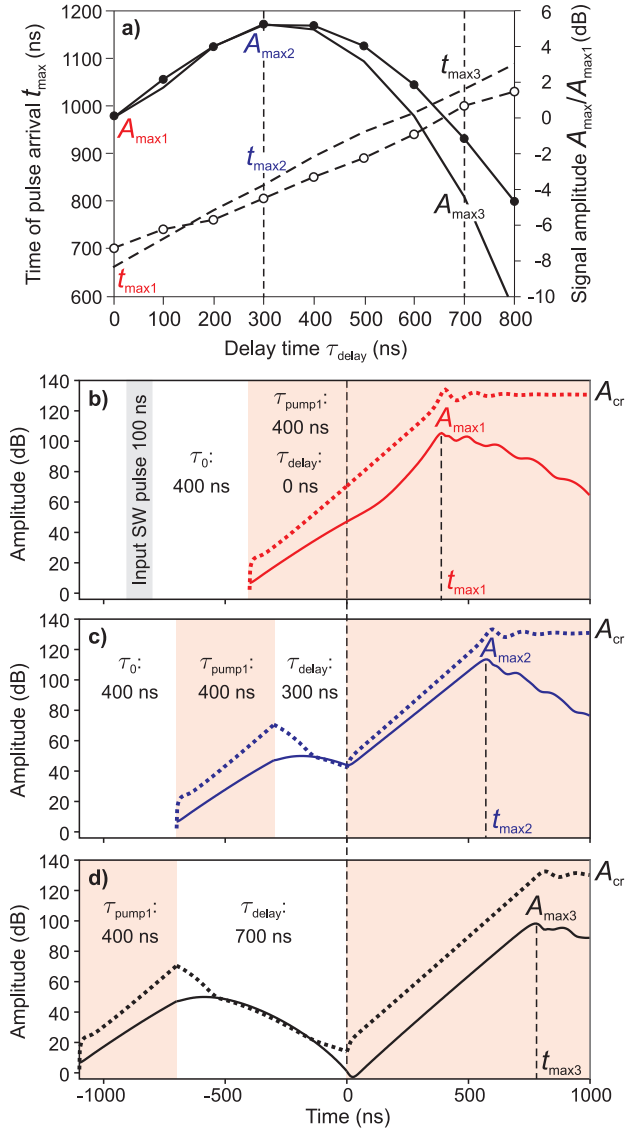}
\caption{ \label{Figure4} (Color online) a) Left axis: measured (circles connected by a dashed line) and simulated (dashed line) time $t_\mathrm{max}$ at which the peak amplitude of the recovered pulse is observed as a function of $\tau_{\mathrm{delay}}$ for $\tau_{\mathrm{pump1}}=400$~ns. Right axis: measured (dots connected by a solid line) and simulated (solid line) peak amplitudes of the recovered pulse for $\tau_{\mathrm{pump1}}=400$~ns.
b-d) Simulated time profiles of the signal (thin solid lines) and of the internal pumping (bold dashed lines). The light red areas indicate the time intervals when the pump is on. The gray area in panel b) indicates the input spin-wave pulse. b) Uninterrupted pumping ($\tau_{\mathrm{delay}}=0$). c) $\tau_{\mathrm{delay}}=300$~ns. d) $\tau_{\mathrm{delay}}=700$~ns. $t=0$ corresponds to switching on the second pump. Duration of the first pump pulse $\tau_{\mathrm{pump1}}=400$~ns.
}
\end{figure}

The structure of obtained equations allows us to renormalize the spin-wave amplitudes $C_{\omega}$ such that $S=1$ and only the ratio $|S/T|$ matters ($T<0$). Initial conditions for the dominant group are random distribution of amplitudes for its frequency components. Initial conditions for the signal group follow from the expression for the frequency spectrum of a rectangular pulse:
\begin{equation}
C_{w}(t_1)=C_0 \exp(i\omega t_1)F(\delta \omega) \frac{\sin (\omega \tau_\mathrm{input}/2)}{\omega \tau_\mathrm{input}/2},
\end{equation}
where $T_\mathrm{tw}$ is the length of the initial traveling wave pulse which generates the standing wave oscillations through the 2-magnon scattering process, and $\tau_0$ is the time interval between the the passage of the traveling wave pulse through the pump area and the begin of the first pump pulse, $C_0$ is some constant, and $F(\delta \omega)$ is equal to one within the dipole gap in the dipole exchange spectrum of travelling spin waves (Fig.~1), where the parametric signal recovery is usually observed ($F(\delta \omega)=1$, for $-\delta \omega/2<\omega<\delta \omega/2$) and vanishes outside the gap ($F(\delta \omega)=0$, for $| \omega | > \delta \omega/2$).

The system of equations (1,2) was solved numerically. We used 200 discrete values of eigenfrequency detuning $\omega$ from the half-pump frequency in the range from $-2\nu$ to $2\nu$. This formed a system of 200 coupled nonlinear equations which was resolved by using 4th-order Runge-Kutta method.

We do numerical calculations for $\tau_{\mathrm{pump1}}=400$~ns and for a number of ratios $|S/T|$, a number of values of rates of parametric gain, and a number of decay rates for both magnon groups. Each time all these parameters, including $C_0$, are chosen such that the restored pulse peaks at the same time as in the experiment when the parametric pumping is not interrupted ($\tau_{\mathrm{delay}}=0$). A natural constraint we impose on the validity of simulation results is a single-peaked shape for the restored pulse for all experimental lengths of the pump-free pause $\tau_{\mathrm{delay}}$.
\vspace{0.5cm}

\textbf{IV Discussion}

A number of simulation runs allows us to locate the area in the parameter space, where the simulated behavior is close to the experimental one displayed in Fig.~3. Figure~4a shows the best fit we obtain. One sees a fair agreement with the experiment of the simulated peak amplitudes of the restored pulse and of the times for the peak arrival $t_\mathrm{max}$ as a function of the pause length $\tau_{\mathrm{delay}}$. This calculation shows that the terms involving $S$ and $T$ coefficients do not contribute to the dynamics during the pump-free pause, as the parametric amplification is quite far from saturation during this time interval. Note that analytical solutions exist for $S=T=0$ and they are in full agreement with our simulations for the first two stages of the considered process ($t<0$ in Fig.~4b-4d).

Figures~4b-4d show the behavior of the macroscopic amplitude of the signal group $A_\mathrm{s} = 10 \log (| \sum_\omega C_{\omega}^\mathrm{s} |^2)$ and of the strength of the internal pumping $A_\mathrm{d} = 10 \log (| \sum_{\omega'} C_{\omega'}^\mathrm{d} |^2)$. One sees that during the pump pause both magnitudes do not vary linearly on the logarithmic scale. The signal-group behavior is close to parabolic on this scale. Note the increase in the amplitude for the signal group during the first 200~ns after switching off the first pump in Fig.~4d. A careful analysis of this stage shows that the increase is due to phase reversal in a parametric-echo-like process \cite{Melkov2001, Herrmann1970}. Recall that the coherence of the originally deterministic signal is lost due to dephasing during the time interval $\tau_0$ between the passage of the traveling wave pulse and the application of the first pump pulse. In the phase reversal process the coherence is restored. The full phase restoration occurs for $\tau_{\mathrm{delay}}=\tau_0$.

The behavior of $A_\mathrm{d}$ is more complicated: for the first 150~ns after the first pumping has been switched off, the internal pump decreases parabolically on the logarithmic scale, then its behavior switches to more or less linear on the same scale. This non-monotonic behavior of the internal pump also originates from dephasing. During the first pump pulse the frequency width of the dominant group reduces significantly due to the frequency selective amplification of the initially thermal signal. The phases of the waves which belong to this group are locked to the external pumping and the behavior of the internal pumping follows the linear behavior of the number of magnons, which is in full agreement with L'vov's S-theory. However, once the first pumping has been switched off, the phase coherence within the group is lost and the magnitude of $A_\mathrm{d}$ starts to decrease. As a result, at large times ($t>300$~ns) the magnitude of  $A_\mathrm{d}$ essentially follows the exponential decrease in the number of magnons, but deviates significantly from this law because of the dephasing for $t<300$~ns.

Both echo-like behavior of the signal group and dephasing in the dominant group contribute to the non-monotonic behavior of the peak amplitude of the restored signal. From Fig.~4 one clearly sees that for $\tau_{\mathrm{delay}}=300$~ns (Fig.~4c) the amplitude of the signal at the end of the pause ($t=0$ in this graph) is quite close to the amplitude of this group at $t=0$ for the uninterrupted pump. However, the internal pump by the dominant group at $t=0$ for $\tau_{\mathrm{delay}}=300$~ns is  smaller than for $\tau_{\mathrm{delay}}=0$ by 25~dB. This allows the recovered signal developing a larger peak amplitude for $\tau_{\mathrm{delay}}=300$~ns than for $\tau_\mathrm{delay}=0$. Obviously, a larger $\tau_{\mathrm{delay}}$ requires a larger time for the restored signal to peak, which results in a linear dependence of $t_\mathrm{max}$ on $\tau_{\mathrm{delay}}$ in Fig.~3a.

In agreement with experiment, the model shows a variation of $A_{\mathrm{max}}$ and $t_{\mathrm{max}}$ with $\tau_{\mathrm{pump1}}$. However, in the model the effect is not so pronounced as in the experiment. We also checked if nonlinear damping in the form of four-magnon scattering processes involving all spin waves existing at the half-pump frequency can contribute to the non-monotonic response. Our calculations for $\tau_{\mathrm{pump1}}=400$~ns have shown that the nonlinear damping is negligible during the pump pause for reasonable values of the nonlinear damping coefficient \cite{Scott2004}. The amplitudes of both signal and dominant groups are just too small during the pause to develop the nonlinear damping. Accounting for the nonlinear damping during the second pump pulse does not lead to qualitative changes in the behavior either, unless one assumes that the coefficient of nonlinear damping is larger than $T$ by several orders of magnitude, which is unreasonable.

As a final note of this section we want to comment on applicability of the two-pulse method for other materials. YIG is a unique material: the linear relaxation time for YIG is 100ns+. This allows dephasing of the wave packet in our experiment well before amplitudes of its constituents drop to the thermal level (see Fig.3, where the typical time scale is hundreds of nanoseconds, which is given by the rate of dephasing.) For comparison, the best metallic magnetic material - Permalloy - has magnetic relaxation time $<5$~ns. This material is operational in the same frequency range as YIG, thus the dephasing rate would be the same, be this experiment conducted with Permalloy. Obviously, for the reason of the much stronger decay rate we would not be able to register any signal recovered by the second pump pulse, as on the time scale of hundred nanoseconds no coherent signal would survive in a Permalloy film to the time of its arrival.
However, the new materials, like Heusler alloys and other half-metals, theoretically may have magnetic losses of the same order as YIG \cite{Kubota, Liu, Trudel}. If quality of these materials is improved in the nearest future, it will be possible to use the two-pulse approach for these materials as well.
\vspace{0.5cm}

\textbf{V Conclusion}

In this work we have investigated the relaxation of a free evolving gas of previously parametrically pumped magnons. The experimental results show a clear deviation from the standard exponential spin-wave decay model. In particular, the inherent magnon damping is found to depend upon the presence of the parametric pumping field. The results are in agreement with the model which accounts for variation of phase coherence for parametrically injected magnon groups during the pump-free pause.
\vspace{0.5cm}

\textbf{Acknowledgment}

Financial support by the DFG within the SFB/TRR49 and by the Australian Research Council is gratefully acknowledged. The authors thank Alexy Karenowska for strong support in the preparation of this paper.


\begin{thebibliography}{24}

\bibitem{Suhl1998}
H.~Suhl, IEEE Trans. Mag. \textbf{34}, 1834 (1998).

\bibitem{Gilbert2004}
T.~Gilbert, IEEE Trans. Mag. \textbf{40}, 3443 (2004).

\bibitem{Azevedo2000}
A.~Azevedo, A.~B.~Oliveira, F.~M.~de~Aguiar, and S.~M.~Rezende, Phys. Rev. B \textbf{62}, 5331 (2000).

\bibitem{Rezende2001}
S.~Rezende, A.~Azevedo, M.~Lucena, and F.~de~Aguiar, Phys. Rev. B \textbf{63}, 214418 (2001).

\bibitem{Dobin2003}
A.~Y.~Dobin and R.~Victora,  Phys. Rev. Lett. \textbf{90}, 167203 (2003).

\bibitem{Gilmore2010}
K.~Gilmore and M.~D.~Stiles, Phys. Rev. B \textbf{81}, 174414 (2010).

\bibitem{Demokritov2006}
S.~O.~Demokritov, V.~E.~Demidov, O.~Dzyapko, G.~A.~Melkov, A.~A.~Serga, B.~Hillebrands, and A.~N.~Slavin,
Nature \textbf{443}, 430 (2006).

\bibitem{Khitun2010}
A.~Khitun, M.~Bao, and K.~L.~Wang, J. Phys. D: Appl. Phys. \textbf{43}, 264005 (2007).

\bibitem{Dziapko}
O.~Dzyapko, V.~E.~Demidov, S.~O.~Demokritov, G.~A.~Melkov, and A.~N.~Slavin, New Jour. Phys. \textbf{9}, 64 (2007).

\bibitem{Suhl1957}
H.~Suhl, J. Phys. Chem. Solids \textbf{1}, 209 (1957).

\bibitem{White1963}
R.~M.~White and M.~Sparks, Phys. Rev. \textbf{130}, 632 (1963).

\bibitem{Serga2007}
A.~A.~Serga, A.~V.~Chumak, A.~Andr\'{e}, G.~A.~Melkov, A.~N.~Slavin, S.~O.~Demokritov, and B.~Hillebrands,
Phys. Rev. Lett. \textbf{99}, 227202 (2007).

\bibitem{Cherepanov1993}
V.~Cherepanov, I.~Kolokolov, and V.~L'vov, Phys. Rep. \textbf{229}, 81 (1993).

\bibitem{Kajiwara2010}
Y.~Kajiwara, K.~Harii, S.~Takahashi, J.~Ohe, K.~Uchida, M.~Mizuguchi, H.~Umezawa, H.~Kawai, K.~Ando, K.~Takanashi, S.~Maekawa, and E.~Saitoh,
Nature, \textbf{464}, 262 (2010).

\bibitem{Uchida2010}
K.~Uchida, J.~Xiao, H.~Adachi, J.~Ohe, S.~Takahashi, J.~Ieda, T.~Ota, Y.~Kajiwara, H.~Umezawa, H.~Kawai, G.~E.~W.~Bauer, S.~Maekawa, and E.~Saitoh,
Nature Mater. \textbf{9}, 894 (2010).

\bibitem{Grishin}
S.~A.~Manuilov, R.~Fors, S.~I.~Khartsev, and A.~M.~Grishin, J. Appl. Phys. \textbf{105}, 033917 (2009).

\bibitem{Ordonez}
O.~V.~Kolokoltsev, C.~L.~Ord\'{o}\~{n}ez-Romero, N.~Qureshi, O.~Cortes-P\'{e}rez, G.~L\'{o}pez-Maldonado, and M.~Avenda\~{n}o-Alejo, Electron. Lett. \textbf{46}, 1387 (2010).

\bibitem{Damon1960}
R.~W.~Damon and J.~R.~Eshbach, J. Appl. Phys. \textbf{31}, 104 (1960).

\bibitem{Kalinikos1996}
B.~A.~Kalinikos, N.~G.~Kovshikov, M.~P.~Kostylev, and H.~Benner, JETP Letters \textbf{64}, 171 (1996).

\bibitem{Kalinikos1997}
B.~A.~Kalinikos and M.~P.~Kostylev, IEEE Trans. Mag. \textbf{33}, 3445 (1997).

\bibitem{Schafer2008}
S.~Sch\"{a}fer, A.~V.~Chumak, A.~A.~Serga, G.~A.~Melkov, and B.~Hillebrands,
Appl. Phys. Lett. \textbf{92}, 162514 (2008).

\bibitem{Chumak2009}
A.~V.~Chumak, A.~A.~Serga, B.~Hillebrands, G.~A.~Melkov, V.~Tiberkevich, and A.~N.~Slavin,
Phys. Rev. B \textbf{79}, 014405 (2009).

\bibitem{L'vov1994}
V.~S.~L'vov, \textit{Wave Turbulence Under Parametric Excitation} (Springer, Berlin, 1994).

\bibitem{Zakharov1975}
V.~E.~Zakharov, V.~S.~L'vov, and S.~S.~Starobinets, Sov. Phys. Uspekhi \textbf{17}, 896 (1975).

\bibitem{Melkov2001}
G.~A.~Melkov, Y.~V.~Kobljanskyj, A.~A.~Serga, V.~S.~Tiberkevich, and A.~N.~Slavin,
Phys. Rev. Lett. \textbf{86}, 4918 (2001).

\bibitem{Herrmann1970}
G.~F.~Herrmann, R.~M.~Hill, and D.~E.~Kaplan, Phys. Rev. B \textbf{2}, 2587 (1970).

\bibitem{Scott2004}
M.~M.~Scott, C.~E.~Patton, M.~P.~Kostylev, and B.~A.~Kalinikos, J. Appl. Phys. \textbf{95}, 6294 (2004).

\bibitem{Kubota} T.~Kubota, S.~Tsunegi, M.~Oogane, S.~Mizukami, T.~Miyazaki, H.~Naganuma, and Y.~Ando,
%Half-metallicity and Gilbert damping constant in Co2FexMn1-xSi Heusler alloys depending on the film composition
Appl. Phys. Lett. \textbf{94}, 122504 (2009).

\bibitem{Liu}
C.~Liu, C.~K.~A.~Mewes, M.~Chshiev, T.~Mewes, and W.~H.~Butler,
% Origin of low Gilbert damping in half metals
Appl. Phys. Lett. \textbf{95}, 022509 (2009).

\bibitem{Trudel}
S.~Trudel, O.~Gaier, J.~Hamrle, and B.~Hillebrands,
%Magnetic anisotropy, exchange and damping in cobalt-based full-Heusler compounds: an experimental review
J. Phys. D: Appl. Phys. \textbf{43}, 193001 (2010).



%%%%%%%%%%%%%%%%%%%%%%%%%%%%%%%%%%%%%%%%%%%%%%%%%%%%%%%%%%%%%%%%%%%%%%%%%%%%%%%%%%%%%%%%%%%%%%%%%%%%%%%%%%

\end{thebibliography}
\end{document}